\documentclass[9pt,twocolumn]{article} %
\usepackage[top=30pt,bottom=40pt,left=48pt,right=46pt]{geometry}

\usepackage{graphicx}
\usepackage{tikz}

\usepackage{amssymb}
\usepackage{amsfonts}
\usepackage{amsmath}
\usepackage{esint}
\usepackage{color}
\usepackage{epstopdf}

\newcommand{\mean}{\textbf{E}}

\newcommand{\pd}[1]{\partial_{#1}}

\newcommand{\kinectTM}{Kinect\texttrademark}
\newcommand{\kinectTMS}{\kinectTM~}

\newcommand{\avgpathS}{\overline\Gamma}

\usepackage{letltxmacro}
\LetLtxMacro{\originaleqref}{\eqref}
\renewcommand{\eqref}{Eq.~\originaleqref}

\begin{document}

\title{Fluctuations around mean walking behaviours\\
in diluted pedestrian flows}

\twocolumn[
  \begin{@twocolumnfalse}

\date{\vspace{-5ex}}
\author{\vspace{-5ex}}
    \maketitle

\centerline{\scshape Alessandro Corbetta }
\smallskip
{\footnotesize
\centerline{Department of Applied Physics,}
\centerline{Eindhoven University of Technology,  The Netherlands}
\centerline{\texttt{a.corbetta@tue.nl}}
\medskip
} %

\smallskip

\centerline{\scshape Chung-min Lee}
\smallskip
{\footnotesize
\centerline{Department of Mathematics and Statistics,}
\centerline{California  State University Long Beach, Long Beach, CA, USA}
\centerline{\texttt{chung-min.lee@csulb.edu}}
\medskip
}
\smallskip

\centerline{\scshape Roberto Benzi}
\smallskip
{\footnotesize
\centerline{Department of Physics and INFN}
\centerline{University of Rome Tor  Vergata, Rome, Italy}
\centerline{\texttt{roberto.benzi@roma2.infn.it}}
\medskip
}
\smallskip

\centerline{\scshape Adrian Muntean}
\smallskip
{\footnotesize
\centerline{Department of Mathematics and Computer Science,}
\centerline{Karlstad University, Karlstad, Sweden}
\centerline{\texttt{adrian.muntean@kau.se}}
\medskip
}
\smallskip

\centerline{\scshape Federico Toschi}
\smallskip
{\footnotesize
\centerline{Department of Applied Physics,}
\centerline{Department of Mathematics of Computer Science,}
\centerline{Eindhoven University of Technology, The Netherlands,}
  \centerline{CNR-IAC, Rome, Italy}
\centerline{\texttt{f.toschi@tue.nl}}

}
\medskip

    \begin{abstract}
      Understanding and modeling the dynamics of pedestrian crowds can help
with designing and increasing the safety of civil facilities.  A key feature
of crowds is its intrinsic stochasticity, appearing even under very diluted
conditions, due to the variability in individual behaviours.
Individual stochasticity becomes even more important under
densely crowded conditions, since it can be nonlinearly magnified and may lead to
potentially dangerous collective behaviours.  To understand
quantitatively crowd stochasticity, we study the real-life dynamics of
a large ensemble of  pedestrians walking undisturbed, and we perform a statistical
analysis of the fully-resolved pedestrian trajectories obtained by a
year-long high-resolution measurement campaign. Our measurements have been carried out 
in a corridor of the Eindhoven University of Technology via a combination of Microsoft
\kinectTMS 3D-range sensor and  automatic head-tracking algorithms.  The temporal homogeneity of our
large database of trajectories allows us to robustly define and 
separate average walking behaviours from fluctuations parallel and
orthogonal  with respect to the average walking path.
Fluctuations include rare events when individuals suddenly change their minds and invert their walking direction. 
Such tendency to invert direction has been poorly studied so far even if it may have important implications on the functioning and safety of facilities.
We propose a novel model for the dynamics of undisturbed pedestrians,  based on stochastic differential equations, 
that provides a good agreement  with our experimental observations, including the occurrence of rare events. 

    \end{abstract}

  \end{@twocolumnfalse}
]

The flow of human crowds is a fascinating scientific topic.  
The interest comes from both its connections with open scientific
challenges related to the development of complex behaviours and pattern formation in non-equilibrium systems~\cite{helbing2001traffic} as well as from its relevance to the design and safety of infrastructures~\cite{fruin1987BOOK}.
Connections with statistical physics~\cite{castellano2009statistical} and   fluid dynamics descriptions~\cite{hughes2003flow} have been used to develop models capable to reproduce some of the features observed in crowds phenomenology~\cite{Mou-pnas,helbing2000simulating,cristiani2014multiscale}. 
From a macroscopic point of view it is no surprise that crowds may be described, at least qualitatively,  by means of fluid-like continuity equations for the local crowd density~\cite{cristiani2014multiscale}.  
 
While it may be tempting to extend this fluid analogy even to the case of rarefied gases and complex fluids as paradigms, respectively, of crowds with low and high pedestrian densities, many more qualitative investigations are needed. A key difference between fluids and crowds is the ``active'' nature of crowd ``particles'' with respect to the ``passive'' nature of particles in ordinary fluids.

Despite the fact that pedestrian crowds are ubiquitous, the
availability of high-quality, high-statistics data is still rather limited. This is
probably related to  technical difficulties in the analysis of camera recordings 
that can be easily affected by varying lighting conditions 
and by the difficulties in the accurate identification of pedestrian positions in images~\cite{DBLP:journals/ijon/BoltesS13}.
When available, high quality data are often limited  to short recordings not allowing an  accurate statistical characterisations of the dynamics. This practically impedes  investigations beyond mean behaviours.
Sufficient statistical accuracy is mandatory to investigate the statistical properties of rare events as the ones, for instance, corresponding to individuals  suddenly changing their direction.

To overcome some of these issues, we have performed a crowd tracking experiment with high space and time accuracy and
with unprecedented statistics. These experimental data allow us to
develop and to validate novel and simple stochastic models capable of
quantitatively reproducing the dynamics of single individual pedestrians as well as of the statistical properties.

\section{Conceptual framework}
The behaviour of single individuals has been modeled in recent literature~\cite{helbing1995social,helbing2000simulating} as being 
subjected to  ``social forces'', geometry constraints (or ``wall forces'') as well as to intrinsic  (random) noise. These models account for both ``voluntary'' as well as ``accidental'' pedestrian motions. 
If such a description is correct, we must observe   non trivial effects which cannot be taken
into account by a purely deterministic dynamics (i.e. by considering only social forces and no noise). 
Indeed, this is exactly what happens. Pedestrians with same starting position and velocity  might exhibit different trajectories, and the random noise in the model should be enough to quantitatively explain this departure.  
Furthermore with a small but  well measurable  probability, some pedestrians abruptly invert their own direction of motion during their walk: the random noise in the model should be able to  reproduce quantitatively such rare events. 

\bigskip

In our experiment sudden inversions of walking direction occur with a probability of one in about thousand pedestrians.  Because of the low frequency of these events, it can be
 challenging  to  study quantitatively and thus explain them
in the context of stochastic mathematical models for single pedestrian behaviour. 
In this paper  we provide evidence, with strong experimental support, 
that such rare events can indeed be explained by the effect of
 ``external'' (nondeterministic) random perturbations.  It is important to underline
that the effect of these rare events can be extremely important in non dilute crowd conditions, as in several situations where crowd disasters occurred (see, e.g.,~\cite{helbing2007dynamics,helbing2012crowd}).

\begin{figure}
\centerline{\def\svgwidth{.3\textwidth}%
\makeatletter{}%
\begingroup%
  \makeatletter%
  \providecommand\color[2][]{%
    \errmessage{(Inkscape) Color is used for the text in Inkscape, but the package 'color.sty' is not loaded}%
    \renewcommand\color[2][]{}%
  }%
  \providecommand\transparent[1]{%
    \errmessage{(Inkscape) Transparency is used (non-zero) for the text in Inkscape, but the package 'transparent.sty' is not loaded}%
    \renewcommand\transparent[1]{}%
  }%
  \providecommand\rotatebox[2]{#2}%
  \ifx\svgwidth\undefined%
    \setlength{\unitlength}{303.62644043bp}%
    \ifx\svgscale\undefined%
      \relax%
    \else%
      \setlength{\unitlength}{\unitlength * \real{\svgscale}}%
    \fi%
  \else%
    \setlength{\unitlength}{\svgwidth}%
  \fi%
  \global\let\svgwidth\undefined%
  \global\let\svgscale\undefined%
  \makeatother%
  \begin{picture}(1,0.62288363)%
    \put(0,0){\includegraphics[width=\unitlength]{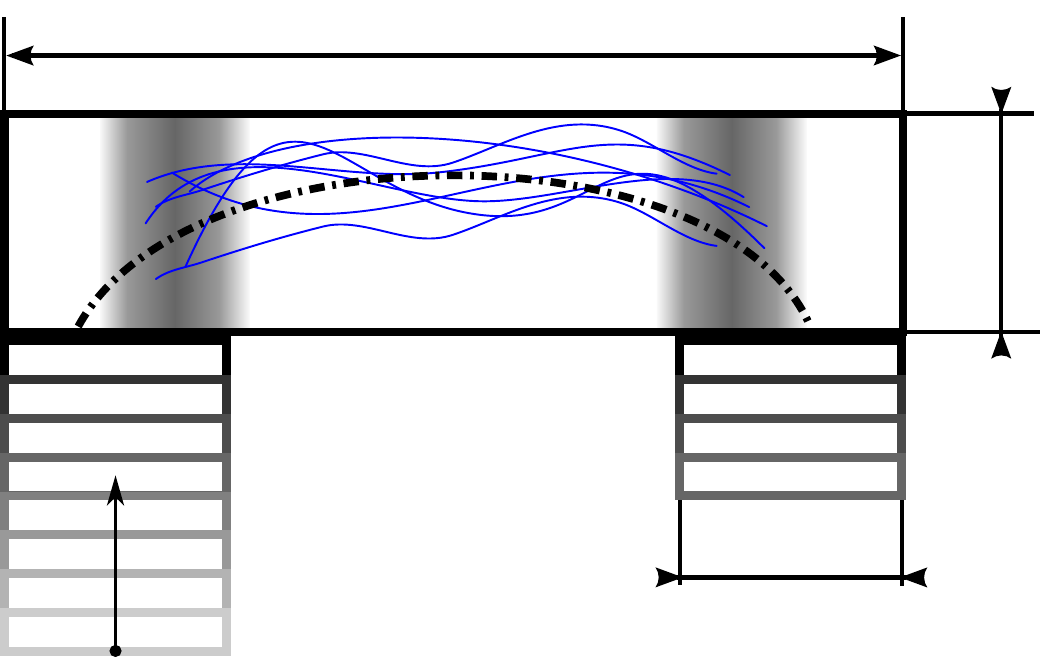}}%
    \put(0.36439875,0.58386454){\color[rgb]{0,0,0}\makebox(0,0)[lb]{\smash{$5.2\,$m}}}%
    \put(0.96098088,0.47947384){\color[rgb]{0,0,0}\rotatebox{-90}{\makebox(0,0)[lb]{\smash{$1.2\,$m}}}}%
    \put(0.68770759,0.01852928){\color[rgb]{0,0,0}\makebox(0,0)[lb]{\smash{$1.2\,$m}}}%
    \put(0.12356245,0.00633827){\color[rgb]{0,0,0}\makebox(0,0)[lb]{\smash{up}}}%
    \put(0.54715474,0.35086997){\color[rgb]{0,0,0}\makebox(0,0)[lb]{\smash{$\overline{\Gamma}(t)$}}}%
    \put(0.72718634,0.44919652){\color[rgb]{0,0,0}\makebox(0,0)[lb]{\smash{$B$}}}%
    \put(0.03403392,0.4460512){\color[rgb]{0,0,0}\makebox(0,0)[lb]{\smash{$A$}}}%
  \end{picture}%
\endgroup%
}
\vspace*{.05in}
\caption{Sketch of the measurement site (staircase landing) with dimensions.  Pedestrians  walk from  region L to 
  R or \textit{vice versa}. From the  individual trajectories (cf. thin lines, only $7$ reported for the sake of readability) we can define an average path $\avgpathS$ (cf. thick dashed line) around which the ensemble of pedestrians fluctuates during their walk.
 \label{fig:schematic}}
\end{figure}
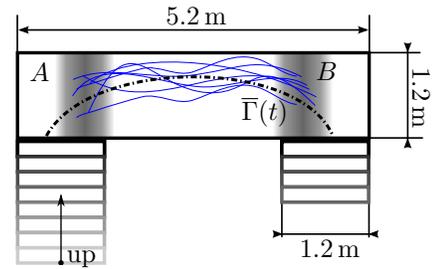

For our purpose, we %
focus on a corridor shaped landing, where the same dynamics  repeats everyday (so that statistics can be arbitrarily increased) and where pedestrians have limited freedom
(they can enter/exit from a restricted
region L and exit/enter from region R).
In our system (sketched in Figure~\ref{fig:schematic})  pedestrians walk subjected to
a very simple geometrical constraint without particular distractions (no pictures,  windows, etc.). The average longitudinal  velocity  is almost the same (within a $10\%$ margin) in the two possible walking directions ($L$ to $R$ and $R$ to $L$).  Let $u$ indicate the longitudinal velocity, we denote by $u_p$ the average value of $u$ (in absolute value).

\begin{figure}
\begin{center}
\centerline{\includegraphics[width=7.7cm]{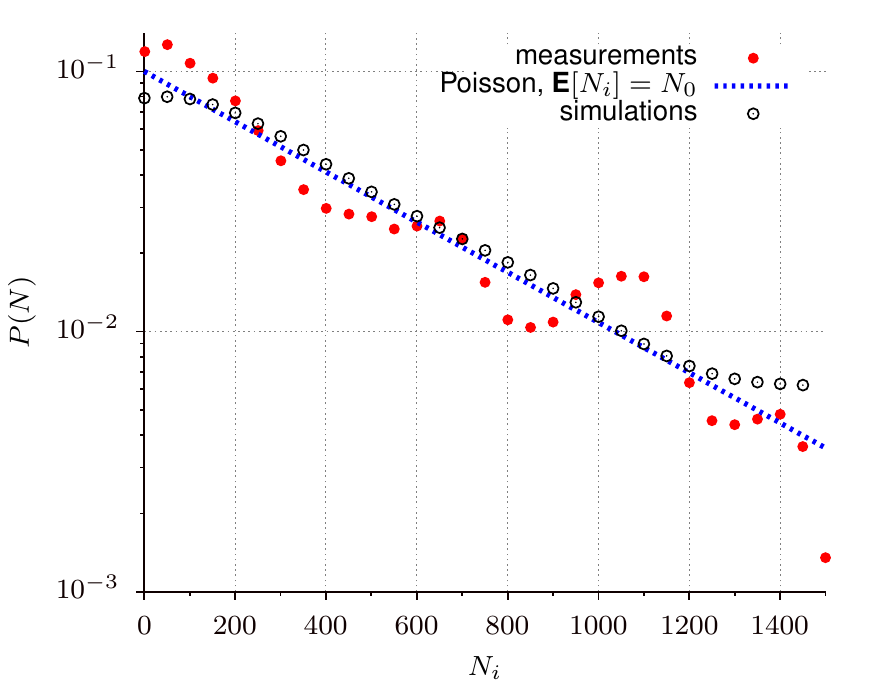}}
\caption{Probability distribution function of the number of pedestrians, $N_i$, passing in the corridor  between two trajectory inversion events (i.e. the number of consecutive crossings of the corridor). Comparison  of measurements (red dots), simulation data from~\eqref{m1}-\eqref{m2} (black open circles) and of a Poisson process with expectation $\mean[N_i] = N_0 = 450$ pedestrians (dotted blue line).}
\label{figturnback}
\end{center}
\end{figure}

Under such conditions, a direction inversion event is simply the change
$u \rightarrow -u$ of the pedestrian's walking direction. The key question is whether the occurrence probability of rare events
 can be {\it quantitatively} related to the amplitude of fluctuations (or {nondeterministic}
 noise if any) as  measured when pedestrians are walking without turning back. 
At first, this idea may appear hopeless because inversion events, as the one we are interested in our case,
 can be due to several subjective {external} factors (e.g. receiving a phone call). %
However, %
 if our  postulation is correct, we should be able to  compute quantitatively
the probability of turning back by a reasonable good measure of the {\it external} stochastic noise. It is the purpose of the present paper to show that this is indeed the case as
shown in Figure~\ref{figturnback}. In Figure~\ref{figturnback}, we report the probability distribution of the number of pedestrians, $N_i$, observed between two consecutive rare events (inversion events). Such probability distribution (red dots) is expected to be exponential since the statistics of rare events follow a  Poisson distribution (after the reasonable assumption that  rare events are independent from each other). The blue dotted line is the best exponential fit of the observations providing $\exp(-N_i/N_0)$ where $N_0 \approx 450$, i.e. on average we observe a rare event every $450$ walking pedestrians. The black open circles are the probability distribution computed using our model (detailed below) and shows a remarkably good agreement with the observations.

\bigskip

In the following we provide the experimental and mathematical details of our approach: first, we give 
the details of our installation, then we describe our stochastic model for pedestrian dynamics, and finally, we compare it against field measurements. %

\begin{figure}
\centerline{\includegraphics[width=.5\textwidth]{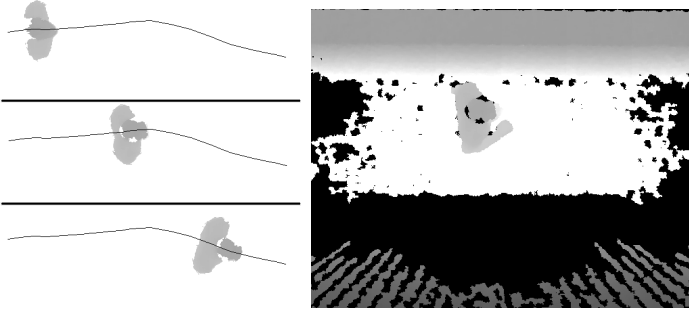}}
\caption{Examples of \kinectTMS depth maps of a single pedestrian walking undisturbed in our measurement site reaching the opposite side (cf. Figure~\ref{fig:schematic}). (a) Three background-less depth maps from three  instants close in time. The reconstructed trajectory of the pedestrian head is superimposed as a solid line. The grey-scale colorization follows the depth levels: darker pixels are closer to the camera plane, thus heads, which are local extrema of the depth field, are darkest. The background, immutable in time, has been subtracted. (b) Example of a raw depth map for the middle frame in (a). Pixels whose depth could not be assessed reliably by the sensor are in black. These typically include far background pixels or shaded regions.
\label{figtrack}}
\end{figure}

\section{Experimental settings}  \label{experiment}
We recorded the trajectories of pedestrians walking in a corridor-shaped landing 
(cf. Figure~\ref{fig:schematic}) in the Metaforum building at Eindhoven University of
Technology (the Netherlands).
Via two staircases at both ends,  the landing connects the 
canteen of the building (ground floor) to the
dining area (first floor). Our installation monitored a  rectangular section in the center
 of the U-shaped walkable area, 
 covering a surface $2.3\,$m long and $1.2\,$m wide (full transversal size). 
Recordings have been carried out on a 24/7 basis for 109  complete working days in the period 
October 2013\,-\,October 2014.

To collect pedestrian trajectories, following~\cite{Seer2014212}, we developed a system with
the following characteristics. 
Via a commercial low-cost Microsoft \kinectTMS 3D-range sensor~\cite{Kinect}
we collect raw overhead depth maps of the corridor (sensor height: $4\,$m; time resolution: $15\,$fps). 
Depth maps are the distance field between observed objects and the sensor plane: such scalar fields can be conveniently encoded in  gray scale pictures (cf. Figure~\ref{figtrack}). \kinectTMS sensors reconstruct depth maps in hardware  (via projection of structured IR light) providing a stream at VGA resolution ($640\,$px$\times 480\,$px). The depth signal enables 
 head detection and hence  the full reconstruction of individual trajectories. 
We report a typical trajectory provided in Figure~\ref{figtrack}. We process the depth map stream offline extracting the head positions frame-by-frame
 (cf.~\cite{Seer2014212}), thus we perform the tracking  in a
Particle Tracking Velocimetry (PTV) fashion~\cite{willneff2003spatio}
via the library OpenPTV~\cite{OpenPTV}. Through this procedure, further described in the SI, we achieve a typical detection and tracking error within a centimeter. 
In particular, head detection reliability is generally high modulo fluctuations due, for instance, to hair or hats ``geometry'', irrelevant for the estimation of trajectories and velocities.

From all pedestrian trajectories connecting L to R and vice versa, we
can define an average path, $\avgpathS$ (sketched in 
Figure~\ref{fig:schematic}, together with few illustrative individual
trajectories). The trajectories of individual pedestrians present some degree
 of stochasticity, and it is thus difficult to disentangle a mean
path from fluctuations at the  single 
trajectory level. Such disentanglement is instead easy and very accurate after ensemble averaging on a large collections of trajectories.
The time resolution of our recordings and the large statistics allow us to achieve a very accurate estimate of the average path $\avgpathS$ (with an error within the millimeter, cf. Figure~\ref{fig:avgpatherr}), enabling us to study the statistics of fluctuations.

\begin{figure*}
\centering
\begin{tikzpicture} %
\node at (0,0){
     \includegraphics[width=5.8cm]{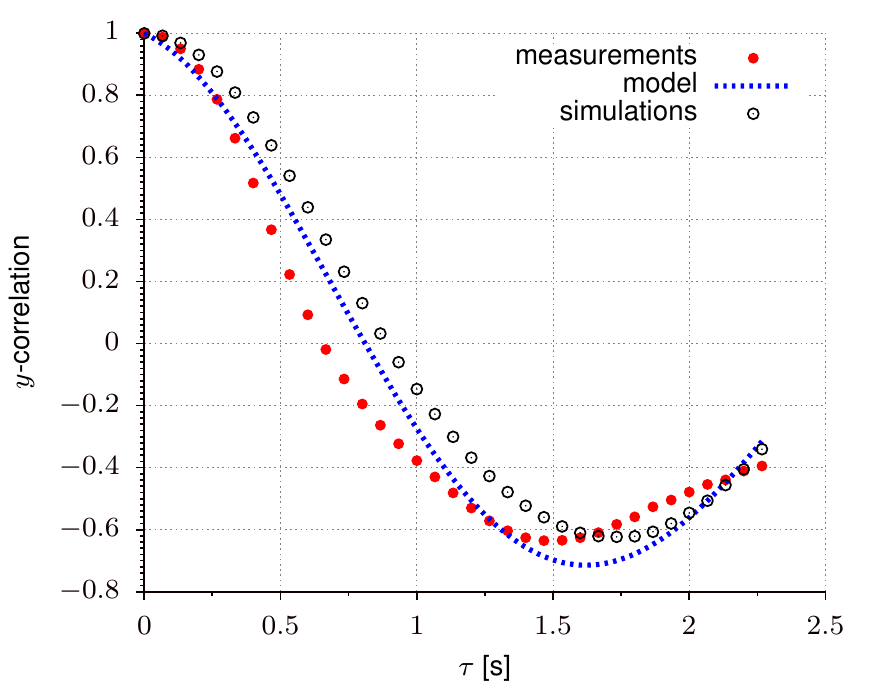}
     \includegraphics[width=5.8cm]{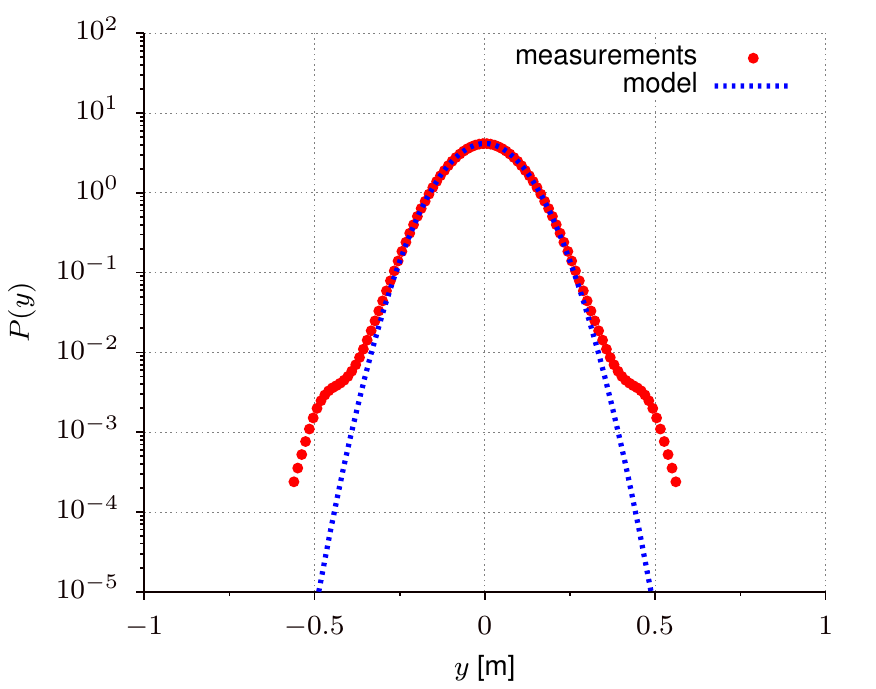}
     \includegraphics[width=5.8cm]{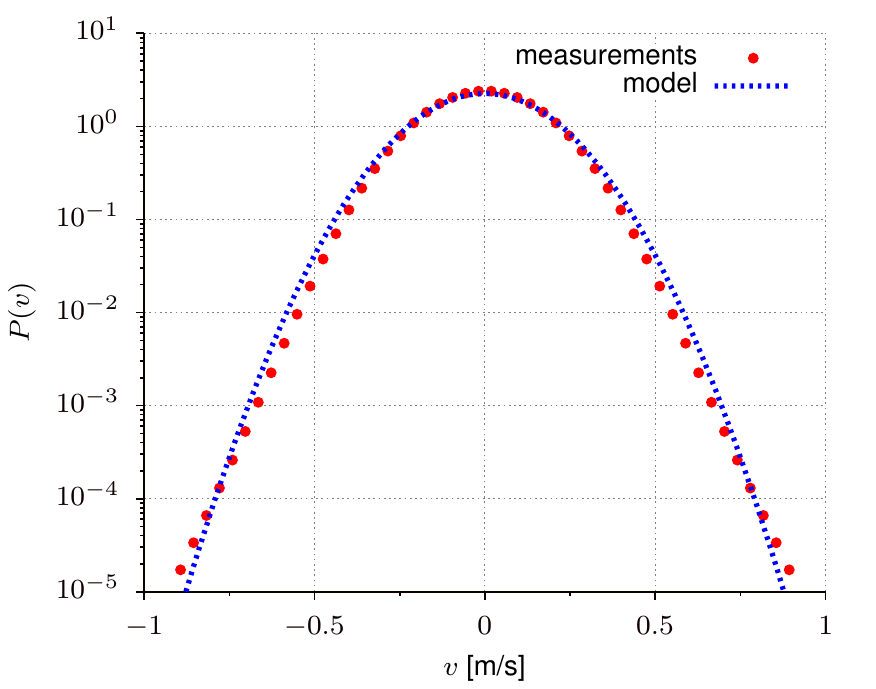}
};
\node at (-7.75,2.25) {\normalsize{(a)}};
\node at (-1.85,2.25) {\normalsize{(b)}};
\node at (+4.05,2.25) {\normalsize{(c)}};
\end{tikzpicture}
\caption{Transversal dynamics: comparison between measurements and model. We model the transversal motion as a harmonically bounded Langevin motion (cf. $y$ and $v$ dynamics in \eqref{m3}-\eqref{m4}). In (a) we report the time-correlation function  of the transversal displacement $y$. The analytic solution (proportional to a cosine function with exponential decay)  is reported as a blue dotted line. Measurements (red dots) and simulations (empty dots) in a domain of equal size are in good agreement with the analytic solution. (b,c) Probability distribution function of, respectively, transversal positions $y$ and transversal velocities $v$. In both cases the analytic solution is a Gaussian distribution (dotted blue line) which is in good agreement with the measurements (red dots). In the case of transversal positions $y$ we observe rare deviations from the Gaussian behaviour at $|y| > 0.4$. These are due to stopping events (cf. peak at $u = 0$ in Figure~\ref{figpotenvx}(c)). We refer the reader to the SI for further details on the calculations.}
\label{figycorr}
\end{figure*}

\begin{figure*}
\centering
\begin{tikzpicture} %
\node at (0,0){
     \includegraphics[width=5.8cm]{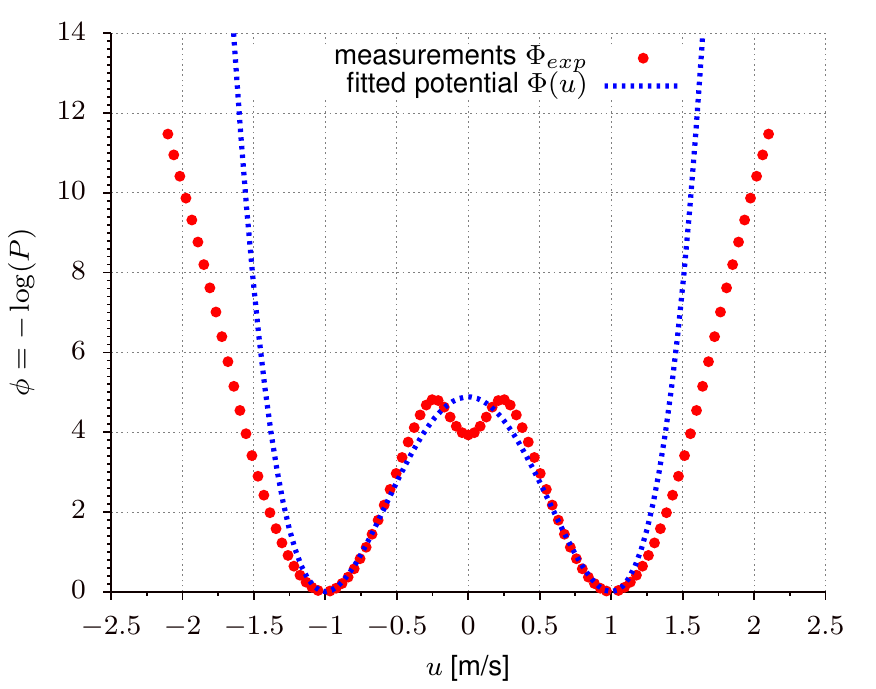}
     \includegraphics[width=5.8cm]{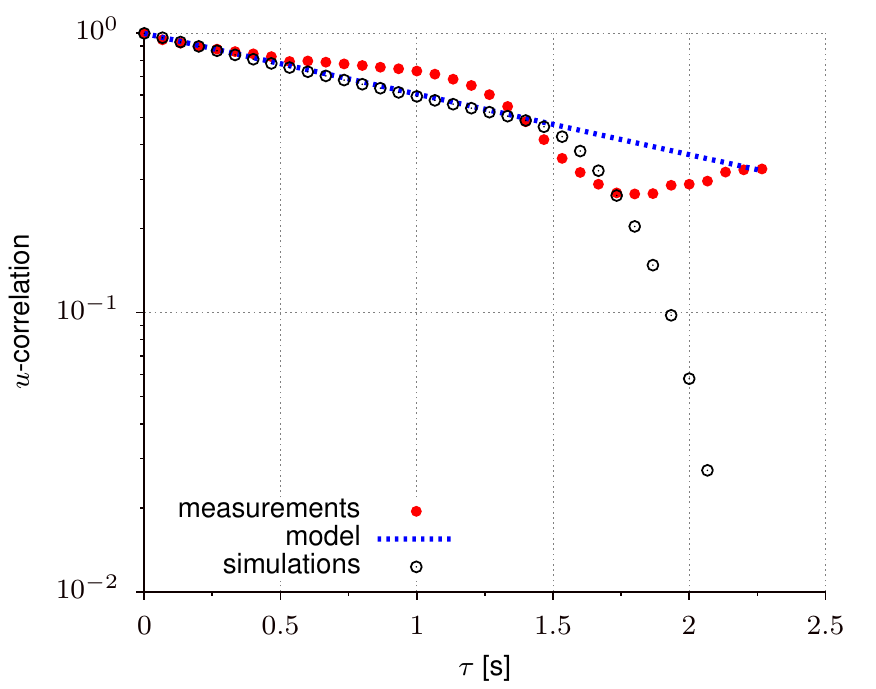}
     \includegraphics[width=5.8cm]{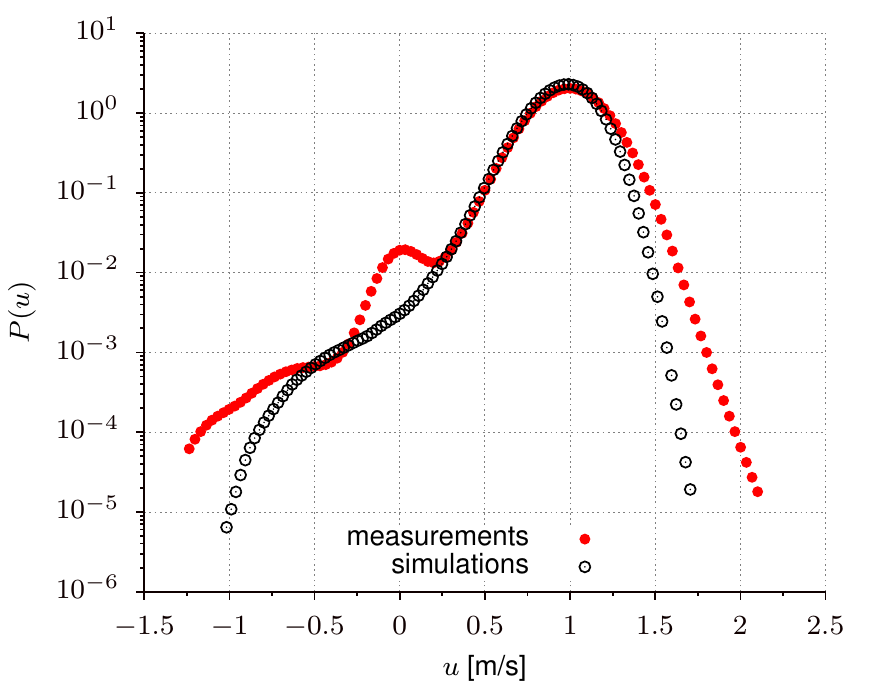}
};
\node at (-7.95,2.25) {\normalsize{(a)}};
\node at (-1.85,2.25) {\normalsize{(b)}};
\node at (+4.05,2.25) {\normalsize{(c)}};
\end{tikzpicture}
\caption{Longitudinal dynamics: comparison between measurements and model. We model the longitudinal motion as a Langevin dynamics in a double well velocity potential (cf. $x$ and $u$ dynamics in \eqref{m1}-\eqref{m2}).
In (a) we compare the experimental potential (after symmetrization of the velocities, cf. \eqref{eq:sym-potential};  red dots) with the rescaled potential $R \phi(u) = R (u^2 - u^2_p)^2$ (dotted line). (b) Time correlation of the longitudinal velocity $u$. The analytic exponential decay of the linearized dynamics ($\exp(-8\alpha u_p^2t)$, cf. \eqref{linearvx}, dotted blue line) is compared with measurements (red dots) and simulations of \eqref{m1}-\eqref{m2} (in a virtual corridor with dimensions similar to those of our experiments; empty dots). The finite size of the corridor is responsible for a deviation from an exponential decay: from simulations, we expect the correlation to decay exponentially  for small times only ($\tau < 1.5\,$s). The measured time correlation (cf. SI for detailed formulas) decays around the expected exponential trend  with larger discrepancies after $\tau > 0.75\,$s. Following the exponential decay at small times we fit the correlation time ($(8\alpha u_p^2)^{-1}$), i.e. $\alpha$. (c) Probability distribution function of longitudinal velocity $u$: comparison between  measurements (red dots) and model (empty dots). The simulated dynamics
captures the entity of the fluctuation as well as the negative velocity tail within the
considered approximation (neglected high velocity behaviour and stops).}
\label{figpotenvx}
\end{figure*}

\section{Dynamics} \label{modeling}

In modelling a single pedestrian walking, our starting point is the introduction of a convenient system of coordinates ($x,y$), where $x$ labels the position in the direction along the corridor and $y$ the transversal position (with $y=0$ corresponding to the center of the corridor). Assuming that there exists no correlation in the longitudinal and transversal dynamics, we model the dynamics in the two directions independently:

\begin{eqnarray}
\label{m1}
\dot{x}(t) &=& u(t) \\
\label{m2}
\dot u(t) &=& f(u) + \sigma_x \dot{W_x} \\
\label{m3}
\dot{y}(t) &=& v(t) \\
\label{m4}
\dot{v}(t) &=& - 2\beta y(t)-2\gamma v(t)+ \sigma_y \dot{W_y}
\end{eqnarray}
where $u$ and $v$ are the velocity components in the longitudinal, $x$, and transversal, $y$, directions and $\beta$ and $\gamma$ are (positive) model parameters.  The structure of the function  $f(u)$ is still to be identified, and the noise terms $\dot{W}_x$ and $\dot{W}_y$ are assumed, for simplicity, independent, $\delta$-correlated in time and Gaussian distributed (these assumptions are conventional, although non mandatory~\cite{romanczuk2012EPJST}). For the time being, we focus on the transversal dynamics where we model the behaviour of a single pedestrian as a linear Langevin equation. There is a priori no reason to believe that a linear approximation is correct or even reasonably good, thus the only way to assess the validity of~\eqref{m3}-\eqref{m4} is to compare the predictions of the model against the outcome of our experiments. 
In Figure~\ref{figycorr} %
we show the $y$ autocorrelation function, the probability density distributions of $y$ and of $v$ respectively. Both $y$ and $v$ show distributions very close to a  Gaussian,  supporting the linear Langevin model in~\eqref{m3}-\eqref{m4}.  The autocorrelation function of $y$ shows good quantitative agreement with the prediction of the linear Langevin equations.  All values of the fitted parameters  are reported in Table~\ref{tabparam}.

\bigskip

Next, we consider the equation for $u$ and we  need thus to identify the function $f(u)$ in~\eqref{m2}. As already pointed out, we have two almost identical velocities characterising the average left-to-right and
right-to-left walk, with an absolute value of about $u_p = 1\,$m/s. Therefore we assume that $f(u_p) = f(-u_p) = 0$, i.e.  the two states $u= \pm u_p$ should correspond to stationary solutions of the deterministic part of~\eqref{m2}. We argue that $u=0$ is also a stationary solution, i.e. $f(0)=0$ and in particular it should be an unstable stationary solution. As we shall see, the assumption on the state $u=0$ is not exactly true and it should be considered as a first approximation. Postponing the question on the state $u=0$, we can reasonably assume that $f(u)$ can be approximated as 
\begin{equation}\label{eq:fu}
f(u) =  - 4\alpha u  (u ^2-u_p^2), 
\end{equation}
where $\alpha$ is a positive parameter that represents the modulating factor of the force.
The above equation is the simplest form of $f(u)$ satisfying our assumptions. Using~\eqref{eq:fu}, we can rewrite~\eqref{m2} in the form:
\begin{equation}
\label{eqvx}
\dot u = - 4 \alpha u (u^2 -u_p^2) + \sigma_x {\dot W_x}.
\end{equation}

Associated with~\eqref{eqvx}, we can consider the stationary probability distribution $P(u)$ given by:
\begin{equation}
\label{theorypdfvx}
P(u)  = {\cal N} \exp \left[ - R \phi(u) \right],
\end{equation}
where $\phi(u) \equiv (u^2-u_p^2)^2$ represents a double-well potential associated with the force $f(u)$, $\cal N$ is a normalisation factor and $R = 2\alpha/\sigma_x^2$.  The way we write $P(u)$ in~\eqref{theorypdfvx} highlights the fact that the stationary probability distribution depends on a single parameter, namely $R$. Note that the probability for a rare event to occur is given by $P(0)/P(u_p) = \exp(-Ru_p^4)$ which corresponds to the well known Kramer's estimate~\cite{berglund2011kramers, kramers1940brownian}. 

To compare our theoretical expectation against experimental data,  we consider the full set of experimental trajectories, in both directions, and we compute the probability density distribution $P_{\exp}(u)$.  From this we construct the potential of the longitudinal dynamics via the relation 
\begin{equation}\label{eq:sym-potential}
\Phi_{\exp} \equiv -\log(\tfrac{1}{2}(P_{\exp}(u) + P_{\exp}(-u))).
\end{equation}
In Figure~\ref{figpotenvx}(a) we compare $\Phi_{\exp}$ to our theory. 
There are two main points to be observed: first, for very large, although rarely occurring, absolute values of $u$, our choice of $f(u)$ is clearly poor; second, at variance with our assumption, the state $u=0$ seems to corresponds to a locally stable state and there exist two unstable states at $u = \pm u_s $ with $u_s \approx 0.2\,$m/s. %
For the second point, what we are missing in our modelling is the relatively small probability to stay at $u=0$ for time longer than the one predicted by~\eqref{eqvx}.  This corresponds to pedestrians stopping walking for a while, possibly taking a phone call.  However, such a time is two order of magnitude shorter than the average transition time  from $u$ to  $-u $. We refrain from increasing the complexity of $f(u)$ to fit the shape of $\Phi_{\exp}$ (though this would easily be possible) since our major goal here is to accurately model the probability of rare trajectory inversion events.  
This goal is relatively simple to achieve, in Figure~\ref{figpotenvx}(a) we chose $R=4.88\,$s$^4$m$^{-4}$ so that the maxima of $R \phi (u) $ corresponds to the two symmetric maxima of $\Phi_{\exp}$. With such a choice, the probability of a rare event, following  Kramer's estimate, is the same in our model and in the experimental data. 

\bigskip

To close our parameter estimation for~\eqref{eqvx}, we need to compute $\alpha$ and/or $\sigma_x$ in an independent way. To this purpose, we consider the case of $u$ close to one of the two ``minima'' shown in Figure~\ref{figpotenvx}(a), say $u=u_p$, and we linearize~\eqref{eqvx} around such a minimum. Upon defining $\delta u=u-u_p$, we can write:
\begin{equation}
\delta {\dot u} = -8 \alpha u_p^2 \delta u + \sigma_x {\dot W_x}.
\label{linearvx}
\end{equation}
From~\eqref{linearvx} %
the correlation function of $\delta u$ should decay as $\exp(-8\alpha u_p^2 t)$ (cf. e.g.~\cite{risken1984fokker}). It is therefore possible to estimate $\alpha$ by computing the correlation function of
$\delta u$ from the experimental data; the results are depicted in Figure~\ref{figpotenvx}(b). 
Although for large time the correlation function does not seem to follow an exponential, at relatively short time we can safely estimate the correlation time as $\alpha \approx 0.0625$\,m$^{-2}$s. Given $\alpha$ we can compute $\sigma_x = \sqrt{2 \alpha/R} \approx 0.16\,$ms$^{-3/2}$. Remarkably, the value of $\sigma_x$ 
is quite close to the value estimated for $\sigma_y$. Although the two noise variances are not constrained to be the same, it is reasonable to argue that the velocity fluctuations should be isotropic, this is in line with what we found. Also, the correlation time $1/(8 \alpha u_p^2) \approx 2\,$s is very close to the correlation time $1/(2\gamma)  \approx 2.4\,$s estimated for the correlation function of $v$. Once more, while there is no reason for the system to be perfectly isotropic, we consider the closeness of the  values of noise variance and correlation times as a non-trivial self-consistency check of our model.

\bigskip

We are now able to accomplish the last and more significant step in our study, namely the analysis of rare inversion events.  To perform a fare comparison between our theoretical approach and the experimental data, we proceed as follows: we simulate numerically~\eqref{m1} and~\eqref{eqvx} with initial condition $x=0$ and $u=u_p$. We integrate the solution up to the point $x=2\,$m (exit) and then we repeat the integration  $N$ times starting with the same initial conditions. Next, we consider the experimental data for the same case, i.e. initial condition $x=0$. The value of $N$ is chosen to be the one obtained in the experiments ($N=72376$). Finally, we compute $P(u)$ as obtained by the numerical simulations and compare it with $P_{\exp}(u)$ from the experimental measurements. Rare events should corresponds to the tail in the probability distribution reaching the state $u = -u_p$. The comparison between the two probability distribution is reported in Figure~\ref{figpotenvx}(c). Although there is a discrepancy at $u=0$  and at extreme values of $u$ (as expected), the overall comparison is extremely good. Figure~\ref{figpotenvx}(c) clearly shows that the probability of rare events, i.e. the individual decision to turn back along the path, can be estimated by the effect of external random perturbations.  This result is apparently in contrast with the intuition that the decision to make an U-turn is an {external} and unpredictable event which cannot be modeled. However, as already pointed out,  it is also possible to consider the shape of function $f(u)$ in~\eqref{m2} and the variance of the noise as a suitable way, in statistical sense, to model this unpredictable individual freedom. We need to stress that our choice of the experimental settings and the very large statistical database are essential for our findings that, to our knowledge, have not been reported by others before. Finally, measurements and simulation are compared in terms of rear events distribution  in Figure~\ref{figturnback} showing very good agreement.

\bigskip

Our result  opens ways to a number of possible investigations. Clearly, in less diluted pedestrians environments, rare events are statistically modified by the effect of other individuals and of their associated  ``social forces''. However, even with due modifications, the possibility of rare inversion events can contribute to non trivial effects, such as the local increase of crowd density. Also, it may be interesting to understand how the probability of rare events is changed by increasing the size of the system (especially in the $x$ direction). For instance, in the case of a longer corridor it is reasonable to expect the emergence of a peak around $u=-u_p$ in the longitudinal velocity distribution in connection to the larger relaxation space allowed to reach stable velocity after inversion. 
We may  also expect, in principle, that our parameters ($\alpha , \beta , \gamma$) are somehow system-size dependent.

\begin{table}
\centering
\label{tabparam}

\begin{tabular}{l r l  | l r l }
$\alpha$ &  $0.0625$ & m$^{-2}$s & $\sigma_x$ &  $0.16$ & ms$^{-3/2}$ \\
$\beta$ & $1.63$ & s/m$^{-2}$ & $\sigma_y$  & $0.16$ & ms$^{-3/2}$\\
$\gamma$ &  $0.207$ & s$^{-1}$ & $u_p$ &   $1.0$ & ms$^{-1}$ \\
\end{tabular}

\caption{Parameters used in the model. $\alpha$: modulating factor of the double-well potential force $f$ governing the longitudinal motion (cf.~\eqref{m2});  $\beta$: stiffness coefficient of the transversal linear Langevin dynamics; $\gamma$:  friction coefficient of the transversal linear Langevin dynamics; $\sigma_x$, $\sigma_y$: white noise intensity in longitudinal and traversal direction; $u_p$: desired mean walking speed.}
 \end{table}

\section{Conclusions}\label{sect:conclusion}
Thanks to an innovative crowd measurement campaign, we investigated quantitatively the statistical properties of single pedestrian dynamics in a simple geometric setting. We reliably measured the  motion of pedestrians in real world conditions for one year long removing several of the constraints and biases of laboratory experiments.  For example, inversion of trajectories would never occur in a laboratory context  where pedestrians are explicitly instructed to walk across a corridor.     
We considered the simplest flow condition possible: \textit{undisturbed} pedestrians walking in a quasi one-dimensional corridor.

Even in this  simple scenario, the dynamics shows different levels of stochasticity consistently and reproducibly  present in the two symmetric cases (left-to-right and right-to-left) considered.  Pedestrians show a randomly fluctuating behaviour around a ``preferred'' average path %
which connects the two extremes of the observed region. Rarely, strongly deviating behaviours, such as long pauses or inversions, are  observed.  %
The presence of such highly deviating behaviours gives  the overall  picture of the dynamics a non-trivial structure, different  from the mean-field average behaviour.

In the same spirit of the statistical analysis of other stochastic systems, we analyse  the dynamics  in terms of probability distribution functions. As a consequence of the extensive measurement campaign performed  we obtained probability distribution functions very well resolved in the tails (extreme events)  and we specifically focused our attention to  positions and velocities pdfs. In the case of the longitudinal velocity, the large deviations measured reflect in a non-Gaussian statistic.

To reproduce such stochastic behaviour and its specific statistical features, we use a Langevin-like equation  with a bi-stable pseudo-potential in the velocity space. 
The stochastic fluctuation of the velocity in the positive velocity well of the potential, excited by a forcing white noise, reflects the natural fluctuations across the preferred path. Furthermore,  the white noise allow us to reproduce  rare transitions responsible of  U-turns, corresponding to transitions from the positive velocity well to the negative well. Remarkably, this behavioural change is not determined a priori, but rather it is the result of a purely random process. 

We believe that the present model can be extended to more complex crowd dynamics like e.g. conditions where the crowd density is high, as common in many civil infrastructures in our cities.

\subsection*{Acknowledgements}
We acknowledge the Brilliant Streets research program of the
Intelligent Lighting Institute at the Eindhoven University of
Technology. AC was partly founded by a Lagrange Ph.D. scholarship granted by
the CRT Foundation, Turin, Italy and by the the Eindhoven University
of Technology, The Netherlands. This work is part of the JSTP research programme ``Vision driven visitor behaviour analysis and crowd management'' with project number 341-10-001, which is financed by the Netherlands Organisation for Scientific Research (NWO).

\bibliographystyle{abbrv}
\bibliography{master}

\subsection*{Supporting Information (SI)}

\subsubsection*{Depth maps acquisition and pedestrian tracking}
Our field measurements are based on the 3D data delivered by an overhead Microsoft \kinectTMS
device~\cite{Kinect}. In addiction to a standard camera, Microsoft a \kinectTMS 
provides a structured-light sensor enabling the evaluation of the depth map of
the filmed scene. Depth maps encode the distance between each point (pixel) in the scene
 and the camera plane. They are typically represented via
gray-scale images (cf. Figure~\ref{figtrack}, darker shades of gray are closer to the camera). %
Following the approach introduced in~\cite{Seer2014212}, and discussed for the current
scenario in~\cite{corbetta2015parameter,AC-high-stat-ped2014}, 
over-head depth maps allow an accurate detection of the
pdestrian positions. Performing an agglomerative  clustering of the foreground part of the depth map through a complete linkage~\cite{duda2012pattern}, we identify pedestrians via a 1:1 correspondence with the clusters appearing in the scene. Clusters are found after cutting the hierarchical clustering dendrogram at an height commensurable with the shoulder size   (cf. reliability analysis in~\cite{Seer2014212}).  Finally, heads are associated
with the ``upper'' part (i.e. having lesser depth) of each cluster ($5^{th}$ percentile). Employing overhead sensors  with vertical top-to-bottom view is not mandatory. In fact, larger recording can be achieved via cameras having pitch angle smaller than $90^o$,  however this comes at the cost of increased probability of mutual pedestrians occlusions and higher automatic detection difficulty. Measurements from sensors in this more general configuration are not treated here. The interested reader can refer e.g. to~\cite{Brscic2013}.

After head positions are assessed on a frame basis, we perform a spatio-temporal matching to reconstruct trajectories.
We employ the tracking algorithms in the  Open Particle Tracking Velocimetry (OpenPTV)
library~\cite{OpenPTV, willneff2003spatio}. We use OpenPTV also to deal with the conversion of 
camera ``pixel'' coordinates to ``metric'' coordinates. 
Calibration has been helped by a ``checker board'' composed of
nine circular holes in a $3\times 3$ configuration (hole diameter: $9\,$cm, hole center distance with first neighbors: $13\,$cm).
This allowed a final resolution of \textit{circa}  $3.9\,$mm per px in
the span-wise direction ($x$) and \textit{circa} $4.1\,$mm per px in the transversal
direction ($y$) around the head plane (approximately $1.7\,$m above the ground).

To reduce noisy fluctuations from 3D reconstruction and head detection, we adopt the Savitsky-Golay smoothing filter~\cite{savitzky1964smoothing}, common in the particle tracking velocimetry community (cf., e.g.,~\cite{PTV-liberzon,luthi2005lagrangian}). We employ a local quadratic approximation based on a symmetric window having width equal to $7$ time samples. 

\subsubsection*{Pedestrian trajectories}
In our continuous recordings, we observed up to six pedestrians walking simultaneously. In this paper we focus on trajectories by individuals moving undisturbed by peers (cf.~\cite{corbettaTGF15,AC-high-stat-ped2014} for an overview of other possible traffic conditions including co-flows and counter-flows). To select these trajectories we operate as follows:
\begin{enumerate}
\item for each trajectory $\gamma$ we compute $L(\gamma)$: the average number of pedestrians observed in the site along this trajectorie. The pedestrian whose trajectory is $\gamma$ is always observed, hence, by construction, $L(\gamma) \ge 1$ holds;
\item we retain all those trajectories $\gamma$ for which $L(\gamma) \leq L_1 = 1 + \epsilon_L$, with $\epsilon_L$ small ($\epsilon_L = 0.05$ in our case). Allowing a small $\epsilon_L$ allows one to include trajectories in which for few frames (in our case typically one) a pedestrian appeared with a peer. 
\end{enumerate}
Relaxing the selection condition  $L_1=1$ enables increased statistics. When $\epsilon_L$ is small we argue a reasonably negligible perturbation on the individual trajectories by the presence of a peer. In fact, at small $\epsilon_L$ two individuals can appear together just when at the opposite sides of the facility one enters and one leaves. 

\begin{figure}
\begin{center}
\centerline{\includegraphics[width=.48\textwidth]{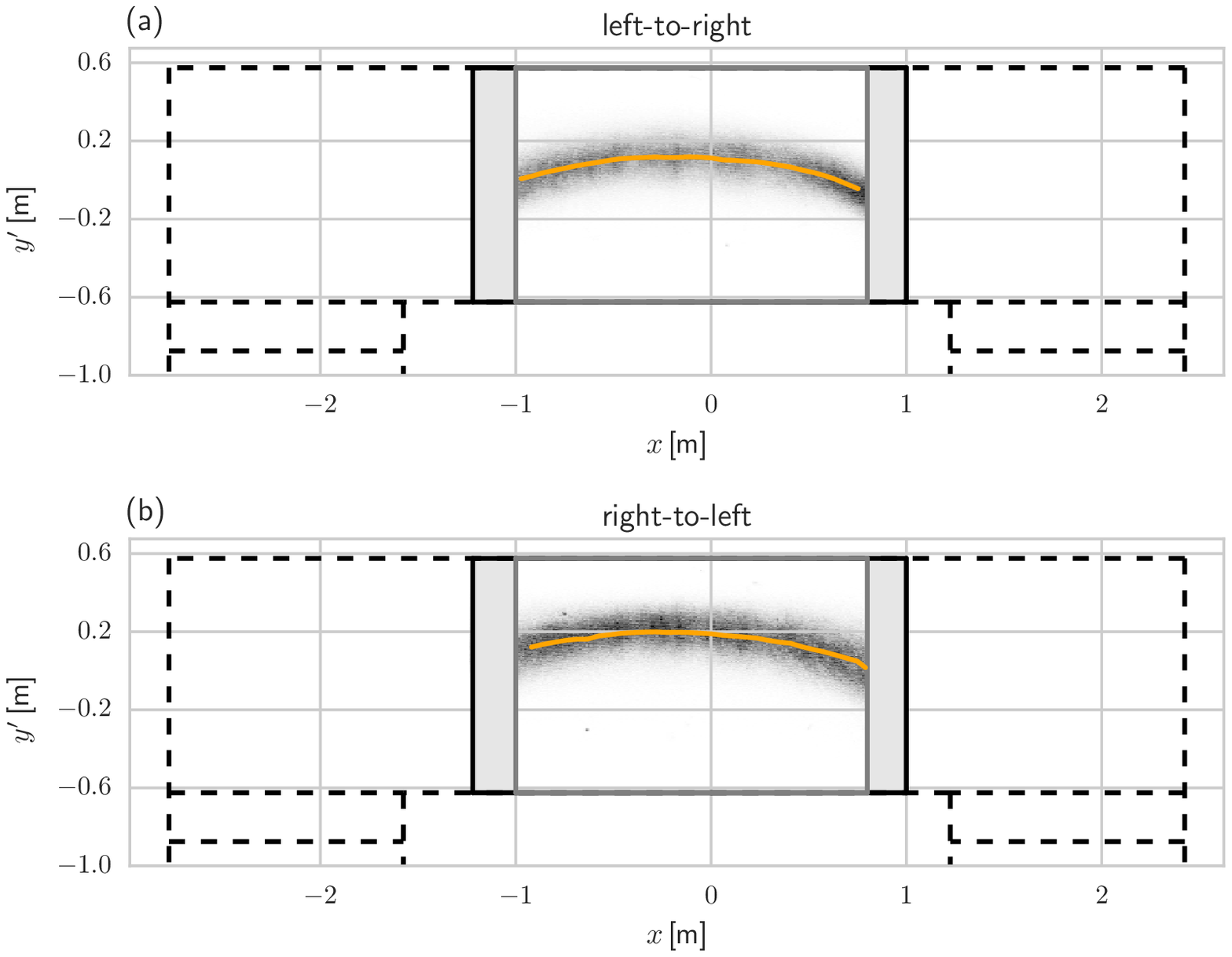}} %
\centerline{\includegraphics[width=.48\textwidth]{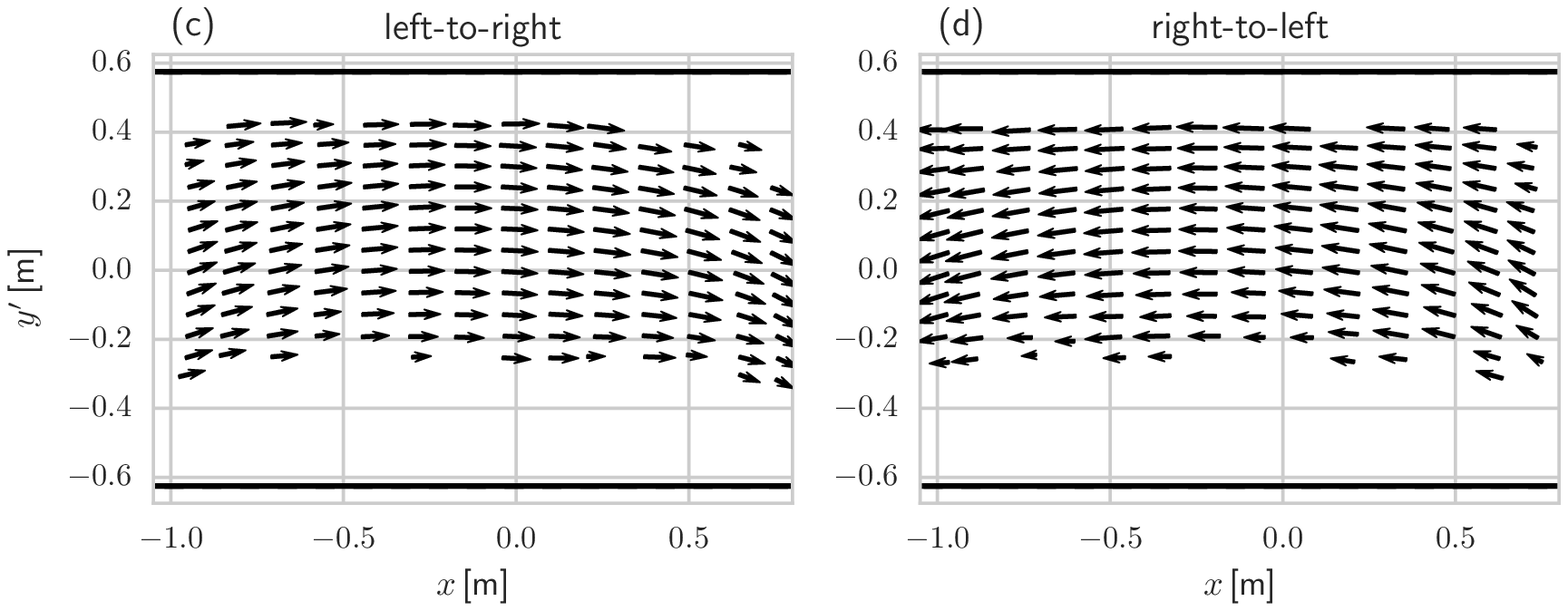}}
\caption{(a,b) Probability density function of pedestrian positions, respectively for pedestrians going left-to-right (a) and right-to-left  (b).  Average paths $\avgpathS$ from \eqref{eq:avg-path} are reported as a solid line. These are calculated after a binning of the measurements in $40$ equal intervals within $[-1.0,0.8]$ in dependence on the $x$ coordinate. The figure axes report Cartesian coordinates aligned with the main directions of the corridor, respectively longitudinal, $x$, and transversal, $y'$ (cf. \eqref{eq:avg-path} and \eqref{eq:walking-y-coord}).  (c,d) Average velocity fields for pedestrian going  left-to-right  (a) and  right-to-left (b). The fields are computed after binning the measurements in a $40\times 40$ grid on the region $[-1.0,0.8]\times[-0.6,0.6]$, and averaging velocity measurements bin-by-bin (fields are downsampled for readability). The evaluation of the transversal walking fluctuation $y$ and the longitudinal and transversal components of the walking velocity $u$ and $v$ (cf. \eqref{m1}-\eqref{m4}) employ the references set in this way. The fluctuation $y$ is the distance (parallel to the $y'$ axis, i.e. \eqref{eq:walking-y-coord}) from the average path (\eqref{eq:avg-path}). Longitudinal and transversal velocity components are computed after a projection of the measured velocity on the (normalized) velocity fields. Components are calculated independently for the two classes of pedestrians, then the contributions are merged to obtain the probability distribution functions in Figures~\ref{figycorr} and~\ref{figpotenvx}.
}
\label{fig:vfield}
\end{center}
\end{figure}

We further employed the following quality checks on the trajectories, to remove faulty or low quality data potentially compromising statistics:
\begin{enumerate}
\item we restrict to fully reconstructed trajectories connecting either of the two virtual boundaries $x_L = -0.8\ m$ and $x_R = 1.0\ m$ (cf. vertical boundary  bands in Figure~\ref{fig:schematic} and in Figure~\ref{fig:vfield}(a,b)) or that feature an unconventionally long time duration (as suggested in~\cite{Brscic2013});
\item selected trajectories are if the order of several tens of thousands. These can still contain detection or tracking errors. We screened them manually, mostly exhaustively, prioritizing trajectories providing outlying values from position or velocity (joint) distributions. Among others, we employed the following empiric trajectory-based quantity. For each trajectory $\gamma$, we define
\begin{equation}
F(\gamma) = \frac{\max_\gamma(s) - \alpha_{0.50,\gamma}(s)}{\sqrt{N}},
\end{equation}  
where, respectively,
\begin{itemize}
\item $s$ is a speed measurement along $\gamma$, i.e. $s = \sqrt{u^2+v^2}$;
\item $\max_\gamma(s)$ denotes the maximum value of $s$ along $\gamma$;
\item $\alpha_{0.50,\gamma}(s)$ denotes the $50^{th}$ percentile (median) of $s$ along $\gamma$;
\item $N$ is the number of samples in $\gamma$.
\end{itemize}
This observable $F$ highlights discrepancies between the maximum and the median speed along a trajectory. Outlying $F$ values are likely synonym of jittery trajectory reconstructions. As large differences between $\max_\gamma(s)$ and $\alpha_{0.50,\gamma}(s)$ are likely to occur in case of trajectories spanning long time intervals, which for our site means one (or more) stop-and-go, we introduce the (empiric) weight $N^{-1/2}$. This weight reduce the $F$ ``penalty'' for long, and possibly correct, trajectories.
\end{enumerate}
We ultimately classify trajectories in dependence on the direction, either left-to-right or right-to-left (with reference to Figure~\ref{fig:schematic}). The classification is performed on the basis of the entering side or, when not possible, considering the average longitudinal velocity. Neglecting differences between the dynamics left-to-right and \textit{vice versa} (cf.~\cite{corbettaTGF15}), we  merge the two classes after reversing the direction of the class right-to-left.

\begin{figure}
\begin{center}
\centerline{\includegraphics[width=.48\textwidth]{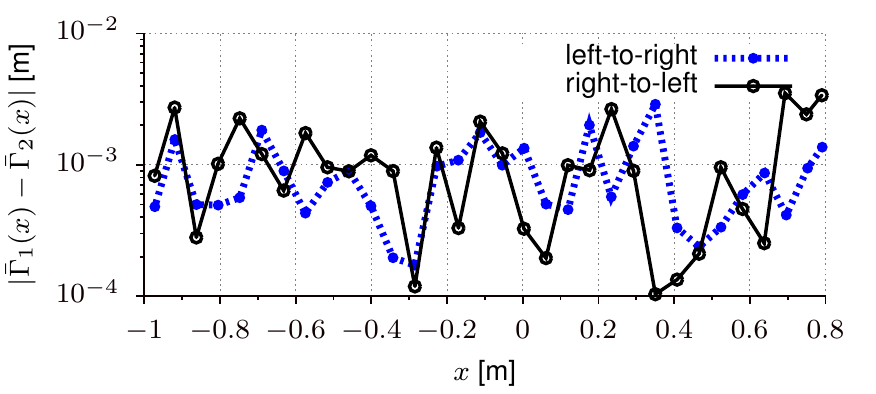}} %
\caption{Error in the evaluation of the average path (\eqref{eq:avg-path}) for pedestrians going  left-to-right  and vice versa. In both cases we split evenly and randomly the measurement sets in two. We report the absolute error on \eqref{eq:avg-path} between the sets, which remains within the millimeter.}
\label{fig:avgpatherr}
\end{center}
\end{figure}

\paragraph{Velocities, positions  and average path}
The U-shape of recording site yields pedestrian trajectories that are slightly curved, as a consequence Cartesian coordinates $x-y'$ that follow the longitudinal and transversal directions of the landing (cf.  Figures~\ref{fig:schematic} and~\ref{fig:vfield}) cannot be used as a reference for the longitudinal and transversal walking direction (coordinates $x-y$ in \eqref{m1}-\eqref{m4}). We define these directions according to curved coordinates following the pedestrian motion, as described in the following. We use adapted coordinate systems obtained independently for the two classes of pedestrians (left-to-right and right-to-left). Thus we merge the components calculated this way to obtain the final probability distributions.

First, to find motion-adapted position coordinates we refer to the average path ($\avgpathS$), that  is curved as the trajectories.  We evaluate average paths from the positions distributions (cf. background in Figure~\ref{fig:vfield}(a,b)). Using a binning in the longitudinal $x$ direction ($40$ bins), we consider per-bin averages of  positions on the $y'$ axis. The average path is given by connecting the bin-dependent $y'$-averages. Using a $x$-dependent parametrization, we  write 
\begin{equation}\label{eq:avg-path}
\avgpathS(x) = (x,\mean[y'|x]), 
\end{equation}
where $\mean[y'|x]$ is the average value of $y'$ for measurements in longitudinal location $x$ (i.e. in the same bin as $x$). Notably, as per the large number of measurements we can assess the average path with low error. For instance, the average distance between the average paths computed splitting our measurements in two random subsets  is about $0.6\,$mm (cf. Figure~\ref{fig:avgpatherr}). Finally, for comparison with the model we remap pedestrians transversal position $y'$ to account for the offset with the average path.  In formulas, a pedestrian in location $(x,y')$ is mapped to location $(x,y)$ where
\begin{equation}\label{eq:walking-y-coord}
y = y' - \mean[y'|x].
\end{equation}
The presence of a preferred path is a key assumption for the dynamics \eqref{m1}-\eqref{m4}. We remark that its physical existence is likely scenario-dependent. For instance, an obstacle in the way may yield two preferred paths, one on each side. On wide corridors preferred paths might be many, up to a continuum.

Second, for the evaluation of the longitudinal and transversal components of pedestrians velocity we refer to the average (Eulerian) velocity field. We consider a two-dimensional spatial binning of our domain composed of $40\times 40$ bins, which define a grid size comparable with the typical head displacement between two following frames (a typical crossing over the observation window takes between $32$ and $37$ frames). We obtain the Eulerian velocity field after an average per bin of all velocity measurements. In Figure~\ref{fig:vfield}(c,d) we report the Eulerian velocity fields for pedestrians going  left-to-right and vice versa. We evaluate the longitudinal  velocity component $u$ by a projection on the local (bin-wise) Eulerian velocity (rescaled to unit modulus). The transversal velocity component $v$ remains defined by difference.

\paragraph{Time correlation}
The time correlation functions for positions are velocity are calculated with respect to the pedestrian state at the domain entrance (initial time-step, $t_0$, of each trajectory). Let $\Xi_t^\gamma$ be the value of observable $\Xi$ (e.g. transversal position or velocity component) that the trajectory $\gamma$ assumes at time $t$. Let $\tilde\Xi_t^\gamma = \Xi_t^\gamma - \mean_t[\Xi_t^\gamma]$ be the fluctuating component of $\Xi$ with respect to the trajectory-wise average $\mean_t[\Xi_t^\gamma]$ at time $t$. The time correlation function of $\Xi$ satisfies
\begin{equation}
C_\Xi(t) = \frac{\mean_\gamma[\tilde\Xi_{t_0}^\gamma\cdot \tilde\Xi_t^\gamma] - \mean_\gamma[\tilde\Xi_{t_0}^\gamma]\cdot \mean_\gamma[\tilde\Xi_{t}^\gamma]}
{
  \sqrt{N(t_0)\cdot N(t)},
}
\end{equation}
where the normalization terms $N(t)$ read
\begin{equation}
N(t) = \mean_\gamma[(\tilde\Xi_t^\gamma - \mean_\gamma[\tilde\Xi_t^\gamma])^2].
\end{equation}

\subsubsection*{Simulations}
We discretize~\eqref{m1}-\eqref{m4} via the two-stage Heun's method (see, e.g.,~\cite{kloeden2011numerical}) using the same data acquisition timestep $\Delta t$, i.e. $\Delta t = 1/15\,$s. Let  $x^n$, $y^n$, $u^n$, $v^n$ approximate the pedestrian state $x(t),y(t),u(t),v(t)$ at instant $t_n = n\Delta t$ (with $n=0,1,2,\ldots,M$), the approximated state at $t_{n+1}$ reads
  \begin{equation}\label{eq:sim1}
    \begin{array}{c c l}
      x^{n+1} &=& x^n+ \frac12 (u^n+u^*)\,\Delta t \\ 
      u^{n+1} &=& u^n - 2\alpha [u^n((u^n)^2-u_p^2) +  u^*((u^*)^2-u_p^2)]\,\Delta t  + \sigma_x\,\Delta \eta \\      
      y^{n+1} &=& y^{n}+\frac12 (v^n+v^*)\,\Delta t\\
      v^{n+1} &=& v^n - \beta (y^n+y^*)\,\Delta t - \gamma (v^n+v^*)\,\Delta t + \sigma_y\,\Delta \eta,
    \end{array}
  \end{equation}
\noindent where
\begin{equation}\label{eq:sim2}
  \begin{array}{c c l}
      x^* &=& x^n +  u^n\, \Delta t \\ 
      u^* &=& u^n - 4\alpha u^n((u^n)^2-u_p^2)\,\Delta t + \sigma_x\, \Delta \eta\\ 
      y^* &=& y^n +  v^n\, \Delta t \\
      v^* &=& v^n -2\beta  y^n\, \Delta t -2\gamma  v^n\, \Delta t + \sigma_y\, \Delta\eta
    \end{array}   
\end{equation}
and $\Delta \eta$ is the integral of a Gaussian white noise in the interval $[t_n,t_{n+1}]$, thus   $\Delta \eta \sim \mbox{Normal}(0,\Delta t)$.
We initialized simulated pedestrians in a virtual corridor at $x=0\,$m, we terminated the advancement of \eqref{eq:sim1}-\eqref{eq:sim2} once one of the two boundaries $x=0\,$m or $x=1.8\,$m was reached. We initialized the transversal position $y$ and  transversal velocity $v$ as zero-averaged normal distributions having the same variance as the experimental measurements.

\subsubsection*{Parameter fitting}
We treat the motion in longitudinal and transversal directions ($x$ and $y$) as independent and so we fit the model parameters. We address here the transversal motion to complement the discussion on the longitudinal motion included in the manuscript.  From \eqref{m2} and \eqref{m4} the probability $P(v,y)$ to observe a given transversal velocity $v$ and position $y$ follows the (stationary) Fokker-Planck equation (e.g.~\cite{GarciaPalacios2004}):
\begin{equation}
  \pd{v}\left\{ (2\beta y +2 \gamma v)P(v,y) + \tfrac{\sigma_y^2}{2}\pd{v} P(v,y) \right\}  - \pd{y}\left\{P(v,y)v\right\} = 0  
\end{equation}
with solution (cf., e.g.,~\cite{pavliotis}):
\begin{equation}
  P(v,y) =  P(v)P(y) =  {\cal N}\exp\left[-\tfrac{2\gamma}{\sigma_y^2} v^2 - \tfrac{4\beta\gamma}{\sigma_y^2}y^2\right]. \label{eq:fokker-planck-sol-transv}
\end{equation}
Values for three parameters, $\gamma$, $\beta$ and $\sigma_y$, are to be identified. We fit the ratios $\tfrac{2\gamma}{\sigma_y^2}$ and $\tfrac{4\beta\gamma}{\sigma_y^2}$ (thus $\beta$) in  \eqref{eq:fokker-planck-sol-transv}  by comparison with the experimental data via the relations
\begin{eqnarray}
  &&-\log[P_{\exp}(v)] \approx \tfrac{2\gamma}{\sigma_y^2} v^2 + K' \\
  &&-\log[P_{\exp}(y)] \approx \tfrac{4\beta\gamma}{\sigma_y^2} y^2 + K'',
\end{eqnarray}
where $P_{\exp}(v)$ and $P_{\exp}(y)$ are, respectively, the empiric distributions of $v$ and of $y$, while  $K'$ and $K''$ are fixed by normalization constraints. We use the time correlation function of $y$ as a third fitting equation. From, e.g.,~\cite{risken1984fokker}, such correlation function satisfies
\begin{equation}
C_y(t) ={\cal N}\exp\left[-\gamma t\right]\left(\cos\omega t + \tfrac{\gamma}{\omega}\sin\omega t\right),
\end{equation}
for the frequency $\omega=\sqrt{2\beta -\gamma^2}$. For the sake of completeness, the (stationary) Fokker-Planck equation associated to  \eqref{m2} and solved by \eqref{theorypdfvx} reads
\begin{equation}
 \pd{u}\left\{ \left( f(u) +\tfrac{\sigma_x^2}{2} \pd{u} \right)P(u)\right\}= 0.
\label{eq:fokker-planck-sol-long}
\end{equation}

\end{document}